\documentclass{elsart}
\usepackage{graphicx}
\usepackage{amssymb}
\begin{document}
\begin{frontmatter}
\title{Cooper pairs in the Borromean nuclei $^6$He and $^{11}$Li using continuum single particle level density}
\author{R. M. Id Betan}
\address{Instituto de F\'isica Rosario (CONICET-UNR), Bv. 27 de Febrero 210 bis, S2000EZP Rosario. Argentina.}
\address{Facultad de Ciencias Exactas, Ingenier\'ia y Agrimensura (UNR), Av. Pellegrini 250, S2000BTP Rosario. Argentina.}
\address{Instituto de Estudios Nucleares y Radiaciones Ionizantes (UNR), Riobamba y Berutti, S2000EKA Rosario. Argentina.}

% \date{\today}
\begin{abstract}
A Borromean nucleus is a bound three-body system which is pairwise unbound because none of the two-body subsystem interactions are strong enough to bind them in pairs. As a consequence, the single-particle spectrum of a neutron in the core of a Borromean nucleus is purely continuum, similarly to the spectrum of a free neutron, but two valence neutrons are bound up in such a core. Most of the usual approaches do not use the true continuum to solve the three-body problem but use a discrete basis, like for example, wave functions in a finite box. In this paper the proper continuum is used to solve the  pairing Hamiltonian in the continuum spectrum of energy by using the single particle level density devoid of the free gas. It is shown that the density defined in this way modulates the pairing in the continuum. The partial-wave occupation probabilities for the Borromean nuclei $^6$He and $^{11}$Li are calculated as a function of the pairing strength. While at the threshold strength the $(s_{1/2})^2$ and $(p_{3/2})^2$ configurations are equally important in $^6$He, the $(s_{1/2})^2$ configuration is the main one in $^{11}$Li. For very small strength the $(s_{1/2})^2$ configuration becomes the dominant in both Borromean nuclei. At the physical strength, the calculated wave function amplitudes show a good agreement with other methods and experimental data which indicates that this simple model grasps the essence of the pairing in the continuum.
\end{abstract}

\begin{keyword}
Continuum \sep Pairing \sep Single particle density \sep Borromean nuclei
\PACS 21.10.Ma \sep 21.60.-n \sep 21.10.Gv
% 04.20.Jb Exact solutions
% 21.10.Gv Nucleon distributions and halo features
% 21.10.Ma Level density
% 21.10.Pc Single-particle levels and strength functions
% 21.30.Fe Forces in hadronic systems and effective interactions
% 21.60.-n Nuclear structure models and methods
% 21.60.Cs Shell model
% 27.20.+n Properties of specific nuclei listed by mass ranges: 6 ≤ A ≤ 19
\end{keyword}
  
\end{frontmatter}

\section{Introduction} \label{sec.introduction}
A Borromean nucleus \cite{1993Zhukov,2004Jensen} is a bound three-body system in which any pair subsystems are unbound. This is so because neither, the bare neutron-neutron interaction nor the core-neutron interaction are strong enough to keep any pair subsystem together. As a consequence, the single-particle spectrum is exclusively continuum. The $^6$He and $^{11}$Li have these characteristics, hence they are both Borromean nuclei, formed by a core plus two neutrons \cite{2002Tilley,2004Tilley}. The properties of these nuclei have been studied in the two-body framework \cite{1987Hansen} as well as using the three-body framework \cite{1997Esbensen,2005Hagino,2014Fortunato,2014Myo,2016Singh} with effective interaction. More elaborate formalisms as the Faddeev equations \cite{1993Zhukov,1998Cobis} and ad initio calculations \cite{2014Redondo,2012Bacca,2009Forssen} has also been used to scrutinize the properties of these exotic nuclei. 

The pairing is a fundamental part of the residual interaction \cite{1964Lane,1987Hansen,2003Dean}. It is particulary important in Borromean nuclei, since the core-neutron system is unbound while the same core wiht two neutrons is bound. Even when the origin of the pairing is unknown, at least two different models provide possible mechanisms for the enhancement of the pairing in the $^{11}$Li nucleus. The first one is through the interplay between the pairing and collective vibration \cite{2001Barranco}. This interpretation is consistent \cite{2010Potel} with the experimental cross section of Ref. \cite{2008Tanihata}. The second mechanism is provided by the tensor force \cite{2007Myo}, which explains the observed Coulomb breakup strength and the charge radius of Ref. \cite{2006Nakamura}.  The simplest pairing interaction is given by the constant pairing \cite{1964Lane}. It will be shown that this effective interaction, even simple, in conjunction with the single particle density, captures the essence of the correlations of the two neutrons in the Borromean nuclei $^6$He and $^{11}$Li.

This paper studies the neutron-neutron pairing correlations in the core of the Borromean nuclei $^6$He and $^{11}$Li. Due to the Borromean character of these two nuclei, the correlations between continuum states are the only one present. Usually these correlations are incorporated through quasibound states obtained by putting the system in a large spherical box. In Refs. \cite{2014Fortunato,2016Singh} the scattering wave functions are used in order to consider explicitly the continuum; in this work instead, the continuum spectrum of energy is handled using the continuum single particle level density (CSPLD). This density is defined as the difference between the mean-field and the free densities \cite{1937Beth,1998Kruppa,2010Ono}.  The used of the CSPLD was implemented earlier in many-body calculations in the Bardeen-Cooper-Schrieffer and Richardson solutions of the pairing interaction in Refs. \cite{2012npaIdBetan,2012prcIdBetan}.

The paper is organized as follows. In section \ref{sec.formalism} the partial wave probability amplitudes in terms of the CSPLD is derived. In section \ref{sec.spr} the single particle representation is defined, while the binding energy and partial wave amplitudes as a function of the strength are calculated in section \ref{sec.results}. The conclusions are given in section \ref{sec.conclusions}. The paper contains an appendix (Appendix \ref{sec.ge}) which gives some details about the CSPLD which modulates the pairing interaction in the continuum.

%%%%%%%%%%%%%%%%%%%%%%%%%%%%%%%%%%%%%%%%%
%%%%%%%%%%%%%%%%%%%%%%%%%%%%%%%%%%%%%%%%%
\section{Formalism} \label{sec.formalism}
The goal of this section is to give the partial wave probability in terms of the partial wave single particle level density. We found it is clearer formulating the problem by starting with continuum discretized wave functions by putting the system in a spherical box (what we call box representation) \cite{1980Maier}. After the equations have been obtained we make the formal limit of the size of the spherical box to infinity. We get a dispersion equation similar to that of Eq. (4) in the two-electrons system \cite{1956Cooper} which includes the continuum single particle level density.

Let us consider the Borromean nucleus as a three-body system formed by an inert core plus two valence neutrons. The Hamiltonian which governs the system reads,
\begin{equation}\label{eq.hbig}
  H= h(1) + h(2) + V
\end{equation}
where $h$ is the single-particle Hamiltonian (see Eq. (\ref{eq.mf})  in sect. \ref{sec.spr}) and $V$ is the residual interaction between the two valence neutrons. The discrete eigenvalues of $h$ are labeled by $\{a,m_a\}=\{ n_a,l_a,j_a,m_a\}$ \cite{1980Lawson}
\begin{equation}\label{eq.hsmall}
  h(\bar{r}) \psi_{am_a}(\bar{r})=\varepsilon_a \psi_{am_a}(\bar{r}) \, ,
\end{equation}
with $\varepsilon_a>0$ for all $a$.

The eigenfunctions of $h$ are used as the single particle representation to build the antisymmetrized and normalized two-neutron bases $|a,b;JM \rangle$. This bases is used two expand the normalized two-neutron wave function $ |\Psi \rangle_{JM}$ in term of the unknown amplitudes $ X^J_{ab}$ \cite{1980Lawson}
\begin{equation}
  |\Psi \rangle_{JM} = \sum_{b\le a} X^J_{ab}  | a,b;JM \rangle 
\end{equation}
with 
\begin{equation}\label{eq.norm}
  \sum_{b\le a} (X^J_{ab})^2  = 1 
\end{equation}

From the Schr\"odinger equation $H |\Psi \rangle_{JM} = E_J |\Psi \rangle_{JM}$ we get the following eigenvalue equation for the three-body problem in the shell model framework
\begin{equation} \label{eq.sec}
  (E_J-\varepsilon_a-\varepsilon_b) X^J_{ab} 
     = \sum_{d\le c} \langle c,d;JM | V | a,b;JM \rangle X^J_{cd}
\end{equation}

We are going to consider as particle-particle effective interaction the constant pairing with matrix elements (m.e.) given by \cite{1964Lane}
\begin{equation}\label{eq.me}
  \langle c,d;JM | V | a,b;JM \rangle 
    = -\frac{G}{2} \sqrt{(2j_c+1)(2j_a+1)} \delta_{J0} \delta_{cd} \delta_{ab} 
\end{equation}
complemented with a partial wave cutoff $l_{max}$ and an energy cutoff $\varepsilon_{max}$ which will be specified in Section \ref{sec.spr}.

From the secular equation (\ref{eq.sec}) and the interaction m.e. (\ref{eq.me}) we get the dispersion relation
\begin{equation}\label{eq.dr}
  1 = \sum_{nlj} \frac{(2j+1)}{2} \frac{G}{2\varepsilon_{nlj}-E_0}
\end{equation}
which gives the $J=0$ correlated two-neutron energies. This expression shows that every state in the discretized continuum, no matter if it represents a resonant or a continuum state \cite{2008IdBetan}, contributes with the same strength to the particle-particle correlation. This is a nonphysical feature since the expectation is that states in resonant configurations will have greater probability to interact with each other and with greater strength that states in continuum configurations. We will see below how this attribute of the constant pairing in the discretized continuum is modified in the continuum representation using single particle level density.

From the secular equation (\ref{eq.sec}) and the normalization condition Eq. (\ref{eq.norm}) we get the two-particle wave function amplitudes
\begin{equation}\label{eq.ampl}
  X^J_{ab} = N \delta_{ab} \frac{\sqrt{2j_a+1}}{2\varepsilon_a - E_0}
\end{equation}
with $N$ the normalization coefficient. Then, the two-particle ground state reads $ |\Psi \rangle = \sum_{nlj} X_{nlj}  | nlj,nlj;00 \rangle $ with
\begin{equation}
  X_{nlj}= N \frac{\sqrt{2j+1}}{2\varepsilon_{nlj} - E_0} \, .
\end{equation} 

We define the partial wave amplitude by summing up the contribution of all positive energy states for each partial wave $(l,j)$ 
\begin{equation}\label{eq.prob}
  X_{lj} = N \sum_n \frac{\sqrt{2j+1}}{2\varepsilon_{nlj} - E_0} \, ,
\end{equation}
where the coefficient $N$ is fixed by the normalization condition $\sum_{lj} X^2_{lj}=1$, and $E_0$ is obtained by solving the dispersion relation (\ref{eq.dr}).

In the Appendix \ref{sec.ge} we show that what makes sense in the limit $R\rightarrow \infty$ of the size of the box, is not  $\lim_{R \rightarrow \infty} f(k_n)$ but $\lim_{R \rightarrow \infty} [f(k_n)-f(k_n^{(0)})]$, i.e, the difference between the correlated and the uncorrelated magnitudes \cite{1937Beth,2010Ono}. This is a practical way to get rid of the density due to the free nucleons. A subtraction prescription like this one was proposed by Bonche et al. \cite{1984Bonche,1998Kruppa,2010Ono} for the calculation of the contribution of unbound states in nuclear Hartree-Fock framework of finite temperature. The probability amplitude $X_{lj}$ in the continuum representation reads,
\begin{equation}
  X_{lj} = \sqrt{2j+1} N \int_0^{\varepsilon_{max}} \frac{g_{lj}(\varepsilon)}{(2\varepsilon - E_0)} d\varepsilon
\end{equation}
with 
\begin{eqnarray}\label{eq.geb}
  g_{lj}(\varepsilon) = \frac{1}{\pi} 
                     \frac{d \delta_{lj}}{d\varepsilon} 
\end{eqnarray}
and $\delta_{lj}(\varepsilon)$ the partial wave phase shift (Appendix \ref{sec.ge}).

The probability amplitude $X_{lj}$ might be negative for some partial wave if $g_{lj}(\varepsilon)$ takes negative values, but the partial probability $X^2_{lj}=(X_{lj})^2$,
\begin{equation}\label{eq.x2lj}
  X^2_{lj} = (2j+1) N^2 \left[ \int_0^{\varepsilon_{max}} \frac{g_{lj}(\varepsilon)}{(2\varepsilon - E_0)} d\varepsilon \right]^2 \, ,
\end{equation}
is a positive magnitude. The value of $N$ is defined such that the normalization $\sum^{l_{max}}_{lj} X^2_{lj}=1$ is fulfilled.

Notice that if we had defined instead, the partial wave probability as $X^2_{lj} = \sum_n  (X_{nlj})^2$, when the limit of the size of the box is taking to infinity, it could happen that $X^2_{lj} \propto\int \frac{g_{lj}(\varepsilon)}{(2\varepsilon - E_0)^2} d\varepsilon$ might be negative if $g_{lj}(\varepsilon)$ takes mainly negative values in the interval $(0,\varepsilon_{max})$.

By taking the box limit of the Eq. (\ref{eq.dr}) we get the following dispersion relation
\begin{eqnarray}
  1 &=& \sum_{lj} ^{l_{max}} \frac{(2j+1)}{2} 
        \int_0^{\varepsilon_{max}}
        \frac{G \, g_{lj}(\varepsilon)}{2\varepsilon-E_0} \, d\varepsilon 
\end{eqnarray}
This expression physically differs with (\ref{eq.dr}) not only because the limit of an infinite box has been taken, but mainly because the density without the free fermion gas is used (see Appendix and Ref. \cite{1937Beth}). Now it is clear that resonant and non-resonant continuum states will not make the same contribution. In the applications it will be shown how the density affects the partial wave occupation probability. We will find that $g_{lj}(\varepsilon)$ is intense around a resonance and small everywhere else. Then, the above expression could be interpreted in the way that the CSPLD modulates the pairing interaction in the continuum.

In terms of the CSPLD $g(\varepsilon)$ we get a dispersion equation similar to that of Eq. (4) in the Cooper's system \cite{1956Cooper},
\begin{equation}\label{eq.dr2}
  \frac{2}{G} = \int_0^{\varepsilon_{max}} 
                \frac{g(\varepsilon)}{2\varepsilon-E_0} \, d\varepsilon  \, .
\end{equation}
with  (Appendix \ref{sec.ge}).
\begin{equation}\label{eq.ge2}
   g(\varepsilon)=\sum_{lj}^{l_{max}} (2j+1) g_{lj}(\varepsilon) \, .
\end{equation}

Equations (\ref{eq.dr2}) and (\ref{eq.x2lj}) give the energy and the probability occupation, respectively, for the two neutrons in Borromean nuclei interacting by a constant pairing force in the continuum representation through the partial wave CSPLD.

%%%%%%%%%%%%%%%%%%%%%%%%%%%%%%%%%%%%%%%%%
%%%%%%%%%%%%%%%%%%%%%%%%%%%%%%%%%%%%%%%%%
\section{Applications}

%%%%%%%%%%%%%%%%%%%%%%%%%%%%%%%%%%%%%%%%%
\subsection{Single particle representation}\label{sec.spr}
The Woods-Saxon (WS) mean field arranges the neutron $0p_{1/2}$ state below the $1s_{1/2}$ state; this order is experimentally found in the nucleus $^5$He but not in $^{10}$Li. The ground state of $^{10}$Li is the state $1/2^+$ which corresponds to the $1s_{1/2}$ state in the shell model picture. In order to reproduce the experimental order in the nucleus $^{10}$Li, it is usual to use parity-dependent parameters for the WS \cite{1997Esbensen,2005Hagino}. Alternatively, we add to the WS a deep Gaussian potential \cite{2005IdBetan} which produces the same effect. The Gaussian parameters are chosen in a way that it strongly (mildly) affects the $s(p)$ state. Doing so, the same mean-field can be use for odd and even states to describe the neutron states in $^{10}$Li.

The single particle Hamiltonian which determines the representation through the eigenfunction of Eq. (\ref{eq.hsmall}) is
\begin{equation}\label{eq.mf}
  h(\bar{r}) = - \frac{\hbar^2}{2\mu} \nabla^2_{\bar{r}}
             + V_{WS}(r)
             + V_g(r)
             + V_{so}(r) (\bar{l} \cdot \bar{s}) \, ,
\end{equation}
where $\bar{r}$ denotes the coordinate of the nucleon and $\mu$ the reduced mass of the core-neutron system. The central and spin-orbit potentials in terms of $r=|\bar{r}|$ are given by the following expressions
\begin{eqnarray}
  V_{WS}(r) &=& - \frac{V_0}{1+exp(\frac{r-R}{a})} \label{eq.mfp1} \\
  V_{so}(r) &=& - \frac{V_{so}}{ra}
                   \frac{2}{\hbar^2} 
                   \frac{exp(\frac{r-R}{a})}{\left[1+exp(\frac{r-R}{a})\right]^2} \label{eq.mfp2} \\
  V_g(r) &=& - V_g e^{-\frac{r^2}{a^2_g}} \label{eq.mfp3}
\end{eqnarray}
with $R=r_0A^{1/3}$. 

The mean-field parameters in (\ref{eq.mfp1})-(\ref{eq.mfp3}) are adjusted using the code Gamow \cite{1982Vertse} in order to reproduce as well as possible the low-lying levels of the nuclei $^5$He and $^{10}$Li. Table \ref{table.mfpar} gives the values of these parameters.
%%%%%%%%%%
\begin{table}[ht] 
\begin{center}
\caption{\label{table.mfpar} Mean-field parameters for a neutron in the $^4$He and $^9$Li mean fields of Eqs. (\ref{eq.mfp1})-(\ref{eq.mfp3}).}
\begin{tabular}{l|l|l}
\hline
 Parameter                            & $^4$He   & $^9$Li      \\
\hline
 $2\mu/\hbar^2$ [MeV$^{-1}$fm$^{-2}$] & 0.038538 & 0.043394  \\
 $a$ [fm]                             & 0.67     & 0.67       \\
 $r_0$ [fm]                           & 1.20     & 1.27       \\
 $V_0$ [MeV]                          & 51.0     & 39.95      \\
 $V_{so}$ [MeV fm$^2$]                & 12.4     & 19.1       \\
 $V_g$ [MeV]                          & 0.       & 609       \\
 $a_g$ [fm]                           & ---      & 0.26        \\
\hline
\end{tabular}
\end{center}
\end{table}

In Table \ref{table.e} whe compare the calculated \cite{1982Vertse} and experimental energies. The real and imaginary complex energies of $^5$He are very similar to the experimental resonant parameters. The ground state energy of $^{10}$Li is real and negative but this nucleus is unbound. This is an antibound state \cite{1994Thompson} with wave number $k_0=-i\, 0.033$ fm$^{-1}$. An antibound state is an unbound single particle state with negative real energy, which would be bound if the mean field were a bit stronger \cite{1986Bohm}. The $p_{1/2}$ energy in $^{10}$Li  was fitted to the average of the two known experimental values.
%%%%%%%%%%
\begin{table}[ht] 
\begin{center}
\caption{\label{table.e} Calculated \cite{1982Vertse} and experimental \cite{nndc,tunl,2002Tilley} low-lying levels (in MeV) of the nuclei $^5$He and $^{10}$Li.}
\begin{tabular}{c|ccc|ccc}
\hline 
                    &&$^5$He&       &                                  & $^{10}$Li&    \\
 State           & Cal.                      && Exp.                    & Cal.    &                   & Exp. \\
\hline
 $0p_{3/2}$ & $0.799-i\, 0.361$&&$0.798-i\, 0.324$
                    & $---$                    &&$---$ \\
 $0p_{1/2}$ & $1.989-i\, 2.310$          &&$2.068-i\, 2.785$
                    & $0.213-i\, 0.053$ &&$0.185\pm 0.040$  \\
                    &                                &&
                    &                                &&$0.240\pm 0.040$ \\
 $1s_{1/2}$ & $---$                       &&$---$                    
                    & $-0.025$   &&$-0.025\pm 0.015$ \\
 $0d_{5/2}$ & $---$                       &&$---$                    
                    & $4.368-i\, 1.670$   &&$---$ \\
\hline
\end{tabular}
\end{center}
\end{table}

The energy cutoff $\varepsilon_{max}$ is defined by using the expression which relates its value with the effective range $r_{nn}=2.75$ fm \cite{1989Slaus} obtained for the three-dimensional delta interaction in ref. \cite{1997Esbensen}
\begin{equation}\label{eq.ec}
  \varepsilon_{max} = \frac{\hbar^2}{m}
             \left( \frac{4}{\pi r_{nn}} \right)^2 \, ,
\end{equation}
which gives $8.884$ MeV using $mc^2=939.57$ MeV and $\hbar c=197.327$ MeV fm. In our calculation we are going to used $\varepsilon_{max}=9$ MeV.

\begin{figure}[ht]
\begin{center}
\vspace{8mm}
\includegraphics[width=0.5\textwidth]{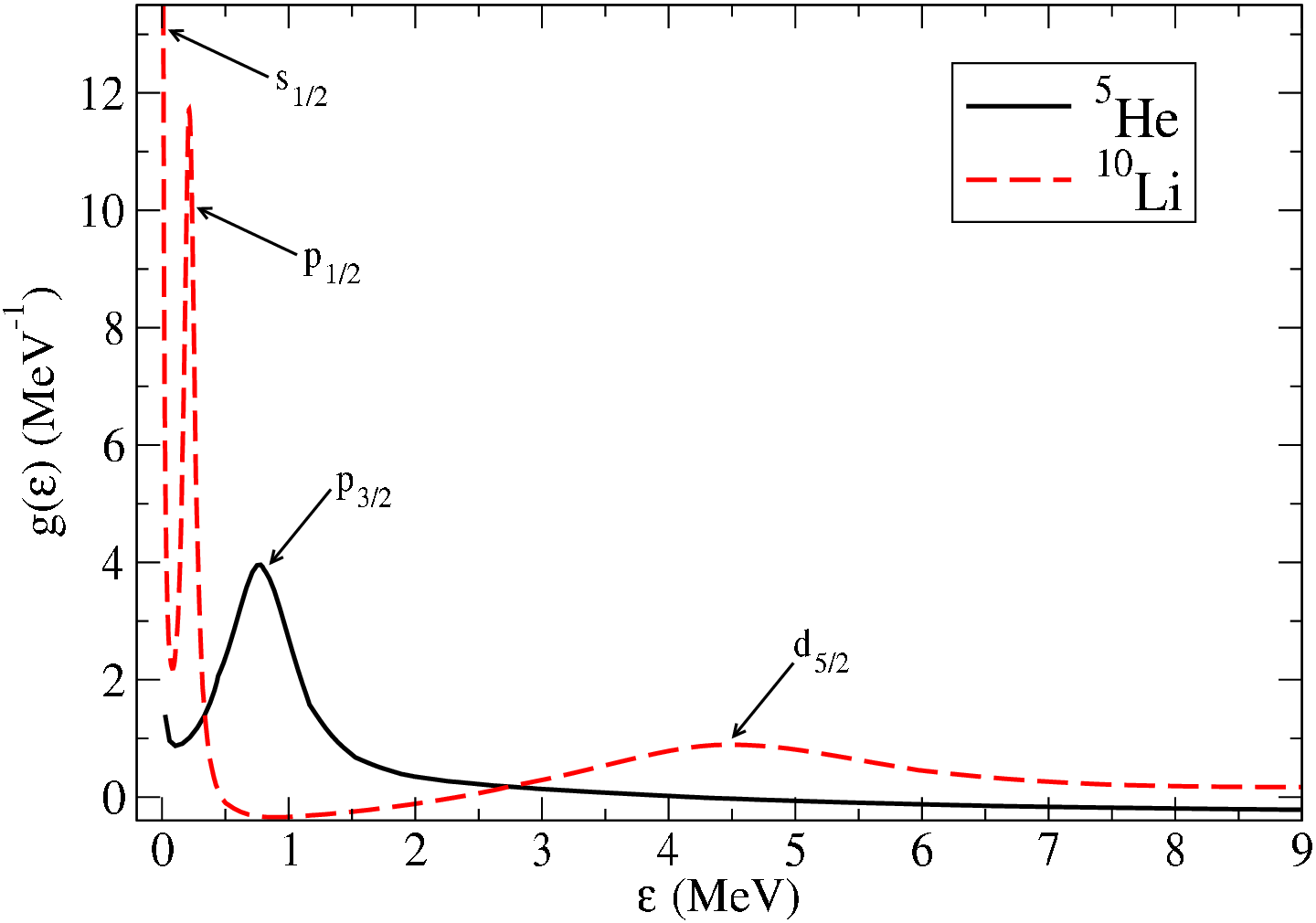}
\caption{\label{fig.ge} (Color online) Neutron continuum single particle level density for $l_{max}=4$ in the $^4$He and $^9$Li cores. The mean fields are defined by the parameters of Table \ref{table.mfpar}.  $(l,j)$ label the main contribution of the partial-wave.}
\end{center}
\end{figure}

Using the mean-field which reproduces the low-lying levels of Table \ref{table.e}, we calculate the neutron partial wave phase shifts $\delta_{lj}(\varepsilon)$ as a function of the energy (up to the energy cutoff $\varepsilon_{max}$) using the code ANTI \cite{1995Ixaru,1996Liotta} and then, we calculate each partial density with Eq. (\ref{eq.geb}). The CSPLD $g(\varepsilon)$ (\ref{eq.ge2}),  shown in Fig. \ref{fig.ge}, is calculated summing up the partial wave densities up to the angular momentum cutoff $l_{max}=4$. Partial wave with bigger angular momentum mildly modify the density for energies $\varepsilon<\varepsilon_{max}$. 

The complex energy poles of the $S$-matrix manifest themselves on the real energy by  modeling the shape of the CSPLD. We found two resonances below $9$ MeV for the $^5$He nucleus (see Table \ref{table.e}). While the $p_{3/2}$ resonance appears as a bump centered around $800$ keV in Fig. \ref{fig.ge}, there is not signal of the $p_{1/2}$ resonance because of its large width ($\sim$ 5 MeV). The finger print of the ground state $1/2^+$ of $^{10}$Li appears as a very sharp structure (labeled as $s_{1/2}$ in Fig. \ref{fig.ge}) in the density, very close to the continuum threshold. This is consistent with the properties of the scattering wave functions $u_{lj}(k,r)$ at low energy \cite{1972Migdal} in the presence of bound or antibound state with energy $\varepsilon \propto k_0^2 \lesssim 0$ close to the threshold 
\begin{equation}
   u_{lj}(kr) \propto \sqrt{ \frac{ 2 k\, | k_0 | }{k^2+|k_0|^2} }\; u_{lj}(|k_0|r) \, ,
\end{equation}
in $^{10}$Li, $| k_0 |=0.033$ fm$^{-1}$. The density has another sharp structure around 200 keV corresponding to the first excited state ($1/2^-$ state). The last structure observed in the $^{10}$Li spectrum is due to the $5/2^+$ resonance around 4.4 MeV.

Notice that the scattering wave functions themselves are not used in the formulation but the CSPLD, i.e. the information of the structure of the continuum is coded in $g(\varepsilon)$ as defined in Ecs. (\ref{eq.geb}) and (\ref{eq.ge2}). The single particle representation is formed by $N_{He}=174$ and $N_{Li}=289$ discretized real energy states for the Helium and Lithium systems, respectively. These numbers of mesh points are chosen so as to make the results stable. The position of the mesh points are selected following the structure of the CSPLD, i.e. states around the resonant energies are favored. The narrower is the resonance, more mesh points are needed to smoothly describe the density; this explains why more states are used to describe the Lithium than Helium up to the same energy cutoff.

%%%%%%%%%%%%%%%%%%%%%%%%%%%%%%%%%%%%%
\subsection{Results} \label{sec.results}
\begin{figure}[ht]
\begin{center}
  \vspace{8mm}
  \includegraphics[width=0.5\textwidth]{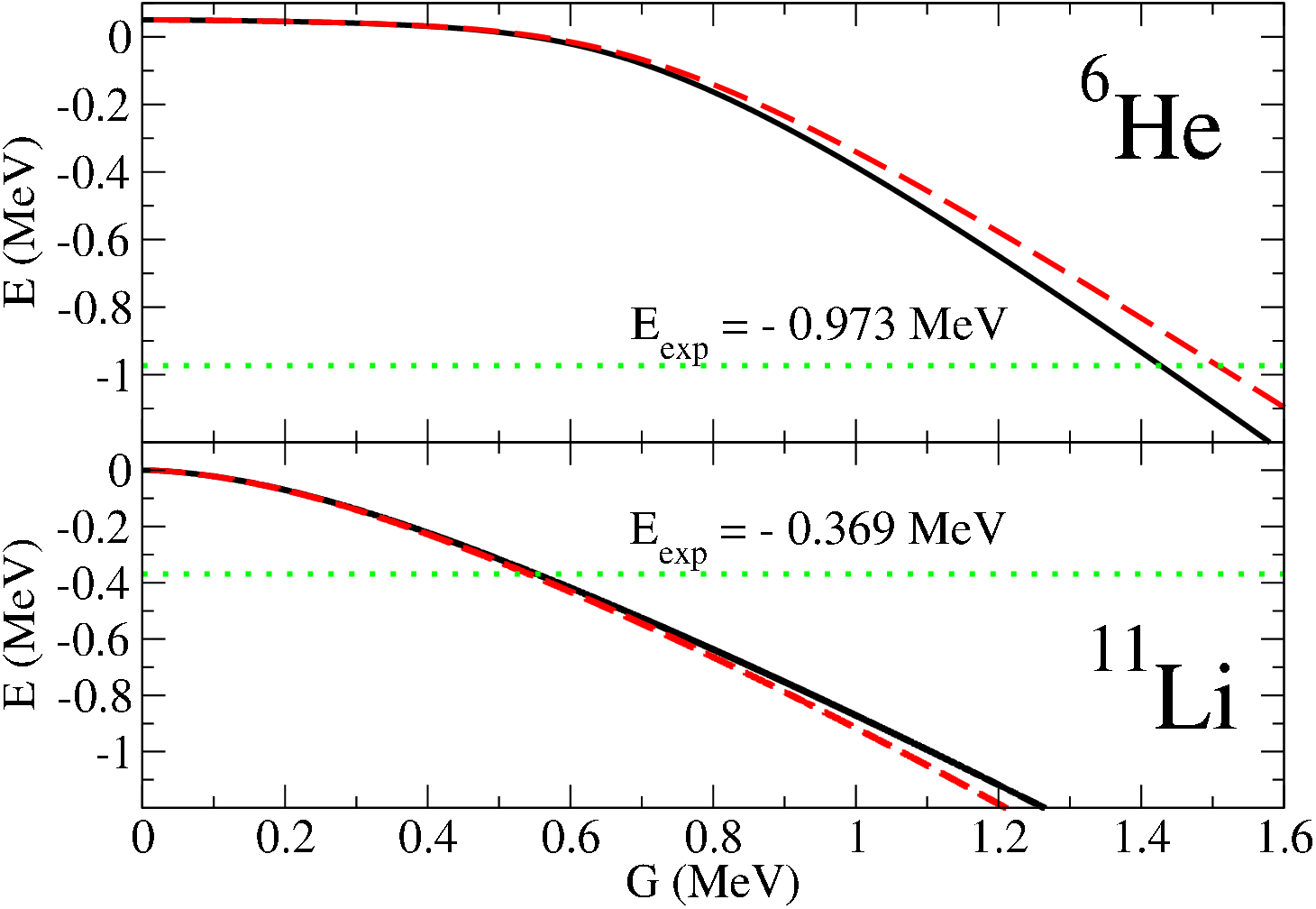}
\caption{\label{fig.EvsgHeLi} (Color online) Ground state energy of the nuclei $^{6}$He and $^{11}$Li as function of the pairing strength $G$ for the energy cutoff  $\varepsilon_{max}=9$ MeV (black solid line) and  $\varepsilon_{max}=18$ MeV (red dashed line). The dotted horizontal lines show the position of the experimental energies from Ref. \cite{2002Tilley} and \cite{2012Wang}, respectively.}
\end{center}
\end{figure}

The ground-state energy of the $^{6}$He and $^{11}$Li nuclei as function of the pairing strength $G$ is calculated and shown in Fig. \ref{fig.EvsgHeLi}. They were obtained from Eq. (\ref{eq.dr2}) using Gauss-Legendre quadrature for the integration. The experimental ground state energies $E_{Exp}=-0.973$ MeV \cite{2002Tilley} and $E_{Exp}=-0.369$ MeV \cite{2012Wang} for $^6$He and $^{11}$Li respectively, are marked by dotted horizontal lines. Let us called \textit{physical strength} $G_{ph}$ the value of $G$ for which the experimental energy is reproduced. For the $^{6}$He system we get $G_{ph}^{He}(9$MeV$)=1.427$ MeV, while for $^{11}$Li we get $G_{ph}^{Li}(9$MeV$)=0.553$ MeV. 

Taking a different energy cutoff $\varepsilon_{max}$ should only renormalized the pairing strength but not change the calculated properties of the system. In order to test this statement, we calculate the wave function amplitudes for a second model space with $\varepsilon_{max}=18$ MeV (using 90 extra mesh points to describe the density in the interval $\varepsilon=(9,18)$ MeV). The evolution of the ground state energy for this second model space is also shown in Fig. \ref{fig.EvsgHeLi}.

For $^6$He, the value of the physical strength for the second model space is $G_{ph}^{He}(18$MeV$)=1.507$ MeV. This figure is larger than for the smaller model space. The usual trend for the strength as a function of the energy cutoff is that the former decreases as the last one increases. The inversion in our model is due to the small negative values of the tail of $^5$He density (see Fig. \ref{fig.ge}). The value of the physical strength for $^{11}$Li is $G_{ph}^{Li}(18$MeV$)=0.542$ Mev. This figure is very similar to the one for the first model space ($\varepsilon_{max}=9$ MeV), indicating that the Lithium system is less sensitive to the energy cutoff than the Helium, probably due to the proximity of $G_{ph}^{Li}$ to the continuum threshold. 

The components $l=0,1$, and $2$ of the occupation probabilities for the two different model spaces (energies cutoff) are given in Table \ref{table.X2HeLi}. Since the results compare well with each other, we will adopt for purpose of comparison with other methods, the model space with  $\varepsilon_{max}=9$ MeV obtained from Eq. (\ref{eq.ec}).
%%%%%%%%%%%%%
\begin{table}
\begin{center}
\caption{\label{table.X2HeLi} Occupation probabilities $X^2_{lj}$ ($l=0,1,2$) for the $^6$He and  $^{11}$Li for the two model spaces at the physical strengths $G_{ph}$.}
\begin{tabular}{c|ccccc}
\hline
 $G_{ph}(MeV)$ & $X^2_{s_{1/2}}$ & $X^2_{p_{3/2}}$ & $X^2_{p_{1/2}}$ & $X^2_{d_{5/2}}$ & $X^2_{d_{3/2}}$ \\
 \hline
   &&& $^6$He && \\
\hline
   1.427     & $0.0853$        & $0.8770$        & $0.0360$        & $0.0015$        & $0.0002$ \\
   1.507   & $0.0920$        & $0.8673$        & $0.0358$        & $0.0043$        & $0.0005$ \\
\hline
   &&& $^{11}$Li && \\
\hline
  0.5527  & $0.3324$        & $0.0070$        & $0.6501$        & $0.0103$        & $0.0002$ \\
 0.5418  & $0.3302$        & $0.0079$        & $0.6508$        & $0.0102$        & $0.0004$ \\
\hline
\end{tabular}
\end{center}
\end{table}

The probability occupation of the ground-state wave function of $^{6}$He is compared with other methods in Table \ref{table.X2He}. The calculated $(p_{3/2})^2$ contribution is not far from the one calculated using density-dependent contact interaction with box basis functions of Refs. \cite{1997Esbensen,2005Hagino}. Our best agreement is with the result obtained using contact-delta interaction in the basis of the continuum scattering $s$, $p$ and $d$ wave functions of Ref. \cite{2016Singh}. Notice that the previous work \cite{2014Fortunato} using only $p$ wave gives much bigger value for the $(p_{3/2})^2$ configuration. The simultaneous comparison of $(p_{3/2})^2$ and $(p_{1/2})^2$ configurations, shows a good agreement with Ref. \cite{2014Myo} which uses Gaussian basis function in the complex scaling framework. A remarkable difference with all other methods is the big contribution of $(s_{1/2})^2$ in our model.
%%%%%%%
\begin{table}
\begin{center}
\caption{\label{table.X2He} Percentage probabilities $X^2_{lj}$(\%) for the main partial-wave components of the ground state wave function of $^{6}$He from different models. The meaning of the abbreviations are: DDCI: density-dependent contact interaction; CI(p): contact interaction within $p$ model space; CxS: complex scaling; ; CI(psd): contact interaction within $psd$ model space.}
\begin{tabular}{c|cccc}
\hline
  Model                           & $X^2_{s_{1/2}}$ & $X^2_{p_{3/2}}$ & $X^2_{p_{1/2}}$ & $X^2_{d}$ \\
\hline
  DDCI \cite{1997Esbensen,2005Hagino}  
                                     & ---        & 83,0  & --- & ---  \\
  CI(p) \cite{2014Fortunato}  
                                     & ---        & 97.2  & 2.8 & ---  \\
  CxS \cite{2014Myo}  
                                     & 0.9       & 91.7  & 4.3 & 3.1  \\
  CI(psd) \cite{2016Singh}  
                                     & 0.8       & 89.7  & 8.0 & 1.4  \\
 This work                      & 8.5     & 87.7  & 3.6 & 0.2          \\
\hline
\end{tabular}
\end{center}
\end{table}

It is experimentally well established that the two main configurations of the two neutrons in the Borromean nucleus $^{11}$Li are $(s_{1/2})^2$ and $(p_{1/2})^2$. Ref. \cite{1997Aoi} shows that the contribution of the second configuration is $(51\pm 6)$\%. In Table \ref{table.X2Li} we compare our result with that of the cluster model of Ref. \cite{1993Zhukov}, the denstiy-dependent contact interaction of Refs. \cite{1997Esbensen,2005Hagino} and that of the Random Phase Approximation (RPA) of Ref. \cite{2001Barranco}. In general we observe a good agreement with all these methods. In particular, the calculated $X^2_s$ contribution is in between the result of the three-body and the collective models while the  $X^2_p$ seems to better agree with the result of Ref. \cite{1997Esbensen}.
%%%%%%%%%%%%%%%%%
\begin{table}
\begin{center}
\caption{\label{table.X2Li} Comparison of the percentage probabilities $X^2_{j}$(\%) configurations for  the ground state wave function of the nucleus $^{11}$Li. The figures for the 'Cluster' model is taken from the average of Table 13 of Ref. \cite{1993Zhukov}.}
\begin{tabular}{c|ccc}
\hline
  Model                                & $X^2_{s}$ & $X^2_{p}$ & $X^2_{d}$ \\
\hline
  Cluster \cite{1993Zhukov} & 39.2  & 54.6 & 1.1         \\
  DDCI \cite{1997Esbensen}  & 23.1  & 61.0  & --- \\
  RPA \cite{2001Barranco}   & 40.0  & 58.0 & 2.0         \\
  DDCI \cite{2005Hagino}     & 22.7  &  59.1 & --- \\  
  This work                          & 33.2 & 65.7  & 1.1         \\
\hline
  Experiment \cite{1997Aoi} &   ---         & 51 $\pm$ 6    & ---         \\  
\hline
\end{tabular}
\end{center}
\end{table}

\begin{figure}[ht]
\begin{center}
  \vspace{8mm}
  \includegraphics[width=0.5\textwidth]{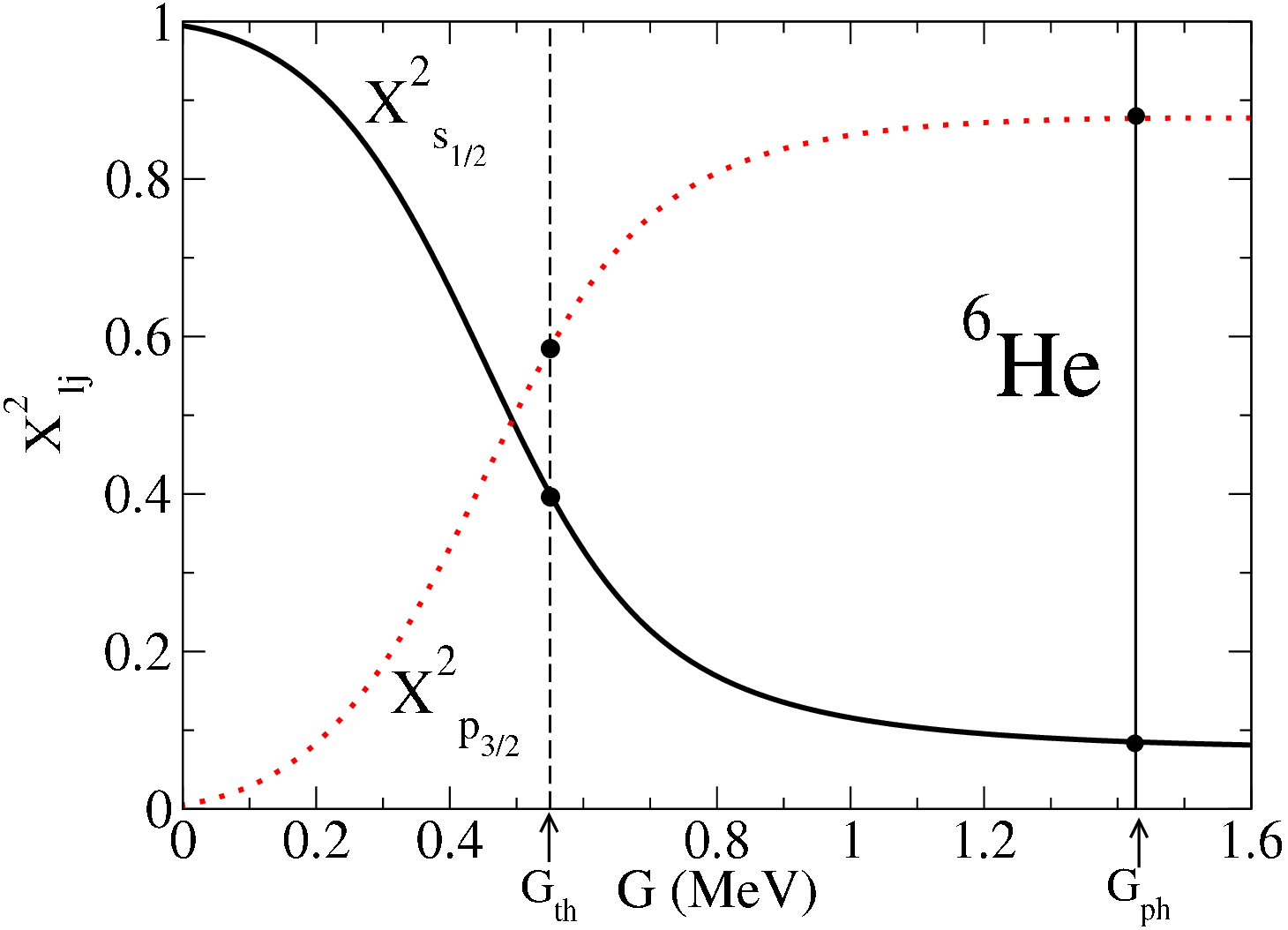}
\caption{\label{fig.X2He} (Color online) The two most important configuration $(lj)^2$ in the $^{6}$He ground state wave function as a function of the pairing strength. The continuum (dashed) vertical line indicates the position of physical (threshold) strength $G_{ph}=1.427$ MeV ($G_{ph}=0.55$ MeV).}
\end{center}
\end{figure}

\begin{figure}[ht]
\begin{center}
  \vspace{8mm}
  \includegraphics[width=0.5\textwidth]{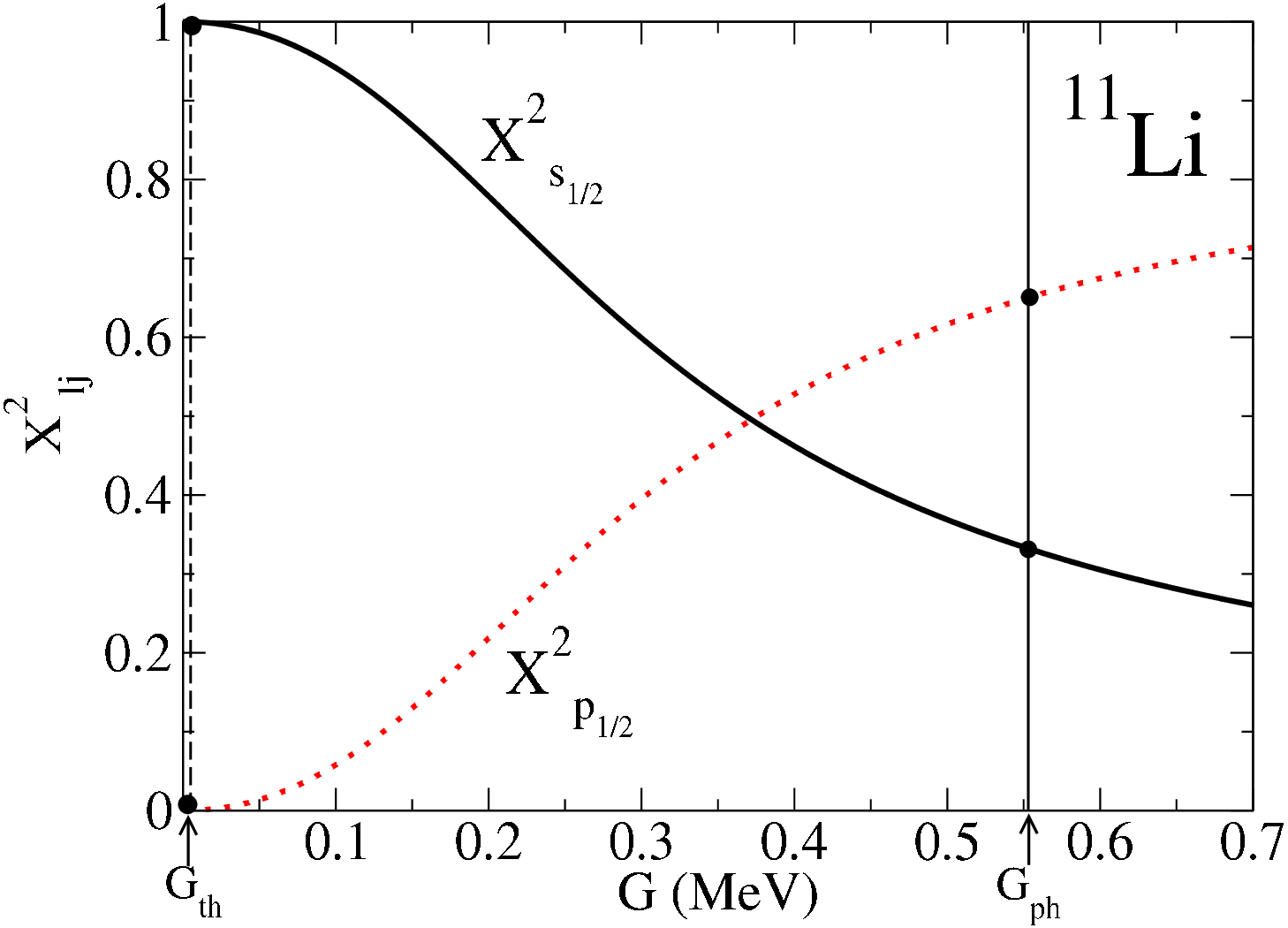}
\caption{\label{fig.X2Li} (Color online) The two most important configuration $(lj)^2$ in the $^{11}$Li ground state wave function as a function of the pairing strength. The continuum (dashed) vertical line indicates the position of physical (threshold) strength $G_{ph}=0.553$ MeV ($G_{ph}=0.005$ MeV).}
\end{center}
\end{figure}

As the last study of the properties of these two Borromean nuclei, we analyze how change the ratio $X^2_{s}$-$X^2_{p}$ as the pairing strength is artificially decreased. Figure \ref{fig.X2He} shows the result for the $^6$He nucleus. At the physical strength, the ground state wave function is dominated by the $(p_{3/2})^2$ configuration. As the strength is decreased, the $p$ contribution reduces is value at the time that the $s$ increases. The system becomes unbound  (changes to positive energy, see Fig. \ref{fig.EvsgHeLi}) for $G_{th} \simeq 0.55$ MeV, called threshold strength (dashed vertical lines Fig. \ref{fig.X2He}). At this value of the strength both configurations $(s_{1/2})^2$ and $(p_{3/2})^2$ are equally important. Figure \ref{fig.X2Li} shows the evolution of the two main components of the wave function in $^{11}$Li. At the physical strength, both $(s_{1/2})^2$ and $(p_{1/2})^2$ configurations are sizable in $^{11}$Li nucleus. As the strength is decreased, the $s$ configuration becomes more and more important, being the dominant one at the threshold strength $G_{th} \simeq 0.005$ MeV. 
%%%%%%%%%%%
%%%%%%%

A common feature of Lithium and Helium nuclei is that small pairing strength favors the configuration $(s)^2$ in detriment of $(p)^2$ in both Borromean nuclei. While a difference between them is that, at the threshold strength the wave function of the Lithium is almost $100$\% $(s_{1/2})^2$, while the two neutrons in the Helium share their strength between the $(s_{1/2})^2$ and $(p_{3/2})^2$ configurations. Figures \ref{fig.X2He} and \ref{fig.X2Li} show that both Borromean nuclei $^6$He and $^{11}$Li are unbound until the pairing force creates enough correlations to unite them all three in a bound system. For the three-body n-n-$^9$Li this transition occurs very close to the continuum threshold, hence a very small correlation between the two neutrons in the $^9$Li core is enough to bind the three-body system. This behavior of the two neutrons in the $^9$Li core resembles very much the behavior of the electrons Cooper pair \cite{1956Cooper,2001Barranco} with the difference, that in our \textit{finite} system, the threshold strength is not zero. The small value of $G_{th}$ may be due to the presence of the antibound state close to the threshold in the n-$^{9}$Li system. The antibound state may also affect other observables, like for example the dipole transition \cite{2009Hagino}.

%%%%%%%%%%%%%%%%%%%%%%%%%%%%%%%%%%%%%%%%%%%%% 
%%%%%%%%%%%%%%%%%%%%%%%%%%%%%%%%%%%%%%%%%%%%% 
\section{Conclusions} \label{sec.conclusions}
The ground state energy and its wave function of the Borromean nuclei $^{6}$He and $^{11}$Li have been studied as a function of the pairing strength using the single particle level density. The model consist of a three-body system with two valence neutrons outside the inert cores $^{4}$He and $^{9}$Li. The neutrons lie in the continuum of their respectively mean fields and they are correlated through a constant pairing interaction modulated by the continuum single particle level density. The single particle representation was defined using the continuum single particle level density defined by the subtraction method \cite{1937Beth}, while  the cutoff energy was settled using the neutron-neutron effective range \cite{1997Esbensen}. In order to compare with other formalism and experimental data, the pairing strength was fixed using the ground state energy of  $^{6}$He and $^{11}$Li. It was found a good agreement with other methods for both nuclei $^6$He and $^{11}$Li. Finally, even when the $(s_{1/2})^2$ configuration becomes dominant as the strength is artificially decreased in both Borromean systems, a seemingly apparent unique feature of the continuum $s$ states in the $^{11}$Li system appears due to the presence of the near-threshold antibound state, i.e. an extremely small (although finite) strength is enough to bind the two-neutron in the $^9$Li core. 
This simple model shows that the essence of the pairing in the continuum is captured through the continuum single particle level density.

%%%%%%%%%%%%%%%%%%%%%%%
\ack
This work was supported  by the National Council of Research PIP-0625 (CONICET, Argentina).

%%%%%%%%%%%%%%%%%%%%%%%%%%%%%%%%%%
%%%%%%%%%%%%%%%%%%%%%%%%%%%%%%%%%%
\appendix
\section{Partial wave single-particle level density}\label{sec.ge}
In this appendix we give details about the continuum single-particle level density (CSPLD) which is use in this work for the constant pairing interaction in the continuum energy representation. This density is derived from the box representation and it is expressed in terms of the derivative of the partial wave phase shift. We closely follow the consideration done by Beth and Uhlenbeck for the calculation of the second virial coefficient \cite{1937Beth}.
 
A partial-wave scattering state is characterized by the angular momentum $l$, the total angular momentum $j$ and the continuum wave number $k$. This generalized eigenfunction of the single-particle Hamiltonian with continuum eigenvalue $\varepsilon=\frac{\hbar^2}{2 \mu} k^2$ (where $\mu$ is the reduced mass) is characterized asymptotically by the phase shift $\delta_{lj}(k)$ \cite{1968Schiff},
\begin{equation}\label{eq.ulj}
  u_{lj}(k,r) \rightarrow
              \sin\left[ k r -l \frac{\pi}{2} + \delta_{lj}(k) \right]
\end{equation}
when $r \rightarrow \infty$.

This asymptotic behavior together with the condition that for a given partial-wave the phase shift tends to zero as $k \rightarrow \infty$, determine $\delta_{lj}(k)$ within a multiple of $\pi$. An increase of the orbital angular momentum makes the single-particle mean-field less important, and for this reason it makes sense to used an orbital angular momentum cutoff $l_{max}$. 

One can discretized the continuum scattering states energy $\varepsilon$ by putting the system into a large spherical box with radius $R$. Then, the box boundary condition, $u_{lj}(k,R)=0$ forces the continuous spectrum to have discrete values $\varepsilon_{nlj}=\frac{\hbar^2}{2 \mu} k_{nlj}^2$. The parameter $n$ denotes the number of nodes (counting the ones at $r=0$) of the function $u_{nlj}(r)=u_{lj}(k_{nlj},r)$ in the interval $[0,R)$. The relation between the number of nodes and the phase shift $\delta_{lj}$ can be obtained through the asymptotic expression (\ref{eq.ulj}) and the boundary condition, given
\begin{equation}\label{eq.nlj}
  k_{nlj} R - l \frac{\pi}{2} + \delta_{lj}(k_{nlj}) = n_{lj} \pi
\end{equation}

If for fixed $\{l,j\}$ one orders the states $\varepsilon_{nlj}$ according to the number of nodes of $u_{nlj}$, then $n_{lj}$ gives the number of levels (without counting the degeneracy) between the bottom of the single particle potential and the energy $\varepsilon_{nlj}$ \cite{1937Beth}. In the limit of the box going to infinity the spectrum $\varepsilon_{nlj}$ becomes continuous and a magnitude like $\sum_n f(k_n)$ changes to \cite{1968Schiff}
\begin{eqnarray}\label{eq.br}
  \lim_{R\rightarrow \infty} \sum_{n} f_{lj}(k_{nlj}) = 
          \int dk \left( \frac{dn_{lj}}{dk} \right) f_{lj}(k) 
\end{eqnarray}
with $\frac{dn_{lj}}{dk}=\lim_{\Delta k \rightarrow 0} \frac{\Delta n_{lj}}{\Delta k}$. Where $\Delta n_{lj} = n_{lj}(k+\Delta k)-n_{lj}(k)$ gives the contribution of all states for which $k$ lies between $k$ and $k+\Delta k$. Using the expression (\ref{eq.nlj}) we get
\begin{eqnarray}
   \frac{dn_{lj}}{dk} &=&
   \frac{1}{\pi}
   \left[
        R + \frac{d \delta_{lj}}{dk}  
   \right]
\end{eqnarray}
 
The summation in (\ref{eq.br}) includes negative-energy bound states and positive-energy discretized continuum states. Single particle energies in the core of Borromean systems are exclusively positive. Then, in the limit $R\rightarrow \infty$ we would have
\begin{eqnarray}
  \lim_{R \rightarrow \infty} 
     \sum_{n\, , \varepsilon_{nlj}>0} f_{lj}(k_{nlj})&=& 
      \int_0^\infty 
      g_{lj}^{(total)}(\varepsilon)
      f_{lj}(\varepsilon)
      d\varepsilon 
\end{eqnarray}
where we introduce the total partial wave energy density
\begin{equation}
  g_{lj}^{(total)}(\varepsilon)
    = \lim_{R \rightarrow \infty}
      \left[
      \sqrt{\frac{\mu}{2\pi^2\hbar^2\varepsilon}}\, R
    + \frac{1}{\pi} \frac{d\delta_{lj}}{d\varepsilon}
      \right]
\end{equation}
The first term, which diverges with the size of the box corresponds to the density of the free nucleon. This can be seen by doing an analogous analysis when the nuclear mean field is zero. In such a case we would have in the passing to the limit,
\begin{eqnarray}
      \lim_{R \rightarrow \infty} 
            \sum_{n\, , \varepsilon^{(0)}_{nlj}>0} f_{lj}(k^{(0)}_{nlj})&=& 
      \int_0^\infty 
      g_{lj}^{(free)}(\varepsilon) 
      f_{lj}(\varepsilon)
      d\varepsilon \nonumber \\
\end{eqnarray}
where $\varepsilon_{nlj}^{(0)}=\frac{\hbar^2}{2 \mu} [k^{(0)}_{nlj}]^2$ are the positive discrete eigenvalues (notice that the condition $\varepsilon^{(0)}_{nlj}>0$ is redundant for the free nucleons in the box) and  
\begin{equation}
   g_{lj}^{(free)}(\varepsilon) =
     \lim_{R \rightarrow \infty} 
               \sqrt{\frac{\mu}{2\pi^2\hbar^2\varepsilon}} \,R
\end{equation}

By taking advantage that $g_{lj}^{(total)}$ and $g_{lj}^{(free)}$ have both the same divergence as a function of the box radius, we used the following recipe in the limit of an infinite box (transition to the continuum) \cite{1937Beth} for a fixed partial wave $(lj)$,
\begin{equation}
 \lim_{R \rightarrow \infty} 
  \left[ 
        \sum_{n\, , \varepsilon_{nlj}>0} f_{lj}(k_{nlj}) 
   -    \sum_{n\, , \varepsilon^{(0)}_{nlj}>0} f_{lj}(k^{(0)}_{nlj})
  \right]
   = \int_0^\infty g_{lj}(\varepsilon) f_{lj}(\varepsilon) d\varepsilon
\end{equation}
where $g_{lj}$ is the partial wave single particle level density with the free nucleons density subtracted
\begin{eqnarray}
  g_{lj}(\varepsilon) = \frac{1}{\pi} 
                     \frac{d \delta_{lj}}{d\varepsilon} 
\end{eqnarray}
i.e., the density so defined is the \textit{change} in the density of single particle states at the energy $\varepsilon$ due to the interaction \cite{2002Carvalho}. With the usual convention $lim_{\epsilon \rightarrow \infty} \delta_{lj}(\epsilon)=0$, the phase shift at zero energy is determined by the Levinson theorem as $\delta_{lj}(0)=n_{lj} \pi$ \cite{1982Newton}. The partial density $g_{lj}(\varepsilon)$ may be either positive or negative depending on the sign of the derivative of the phase shift. For example if for a specific $\{ lj \}$ there are no resonance and $n_{lj} \ne 0$, the phase shift will decrease monotonically from $n_{lj} \pi$ to zero \cite{1982Newton} and the partial CSPLD will be negative for all values of the energies up to infinity. The draw-back of this ``density'' to be negative is compensated by the fact that it gives the structure of the continuum, i.e. for resonant partial wave $g_{lj}(\varepsilon)$ is positive around the resonant energy and its amplitude much bigger than for non-resonant partial waves.

The continuum single particle level density (CSPLD) results from the sum of each partial wave CSPLD $g_{lj}$,
\begin{eqnarray}\label{eq.ge}
  g(\varepsilon) &=& \sum_{lj} (2j+1) g_{lj}(\varepsilon)
\end{eqnarray}

%-------------
% Bibliography% 
%\bibliographystyle{elsarticle-num} 
%\bibliography{richardson}{}

\begin{thebibliography}{10}
\expandafter\ifx\csname url\endcsname\relax
  \def\url#1{\texttt{#1}}\fi
\expandafter\ifx\csname urlprefix\endcsname\relax\def\urlprefix{URL }\fi
\expandafter\ifx\csname href\endcsname\relax
  \def\href#1#2{#2} \def\path#1{#1}\fi

\bibitem{1993Zhukov}
M.~V. Zhukov, B.~V. Danilin, D.~V. Fedorov, I.~J. Bang, J. M.~Thompson, J.~S.
  Vaagen, Phys. Rep. 231 (1993) 151.

\bibitem{2004Jensen}
A.~S. Jensen, K.~Riisager, D.~V. Fedorov, E.~Garrido, Rev. Mod. Phys. 76 (2004)
  215.

\bibitem{2002Tilley}
D.~R. Tilley, C.~M. Cheves, J.~L. Godwin, G.~M. Hale, H.~M. Kofmann, K.~J. H.,
  C.~G. Sheu, H.~R. Weller, Nucl. Phys. A 708 (2002) 3.

\bibitem{2004Tilley}
D.~R. Tilley, K.~J. H., J.~L. Godwin, D.~J. Millener, J.~Purcell, C.~G. Sheu,
  H.~R. Weller, Nucl. Phys. A 745 (2004) 155.

\bibitem{1987Hansen}
P.~G. Hansen, B.~Jonson, Europhys. Lett. 4 (1987) 409.

\bibitem{1997Esbensen}
H.~Esbensen, G.~F. Bertsch, K.~Hencken, Phys. Rev. C 56 (1997) 3054.

\bibitem{2005Hagino}
K.~Hagino, H.~Sagawa, Phys. Rev. C 72 (2005) 044321.

\bibitem{2014Fortunato}
L.~Fortunato, R.~Chatterjee, J.~Singh, A.~Vitturi, Phys. Rev. C 90 (2014)
  064301.

\bibitem{2014Myo}
T.~Myo, Y.~Kikuchi, H.~Masui, K.~Kat\={o}, Progress in Particle and Nuclear
  Physics 79 (2014) 1.

\bibitem{2016Singh}
J.~Singh, L.~Fortunato, A.~Vitturi, R.~Chatterjee, Eur. Phys. J. A 52 (2016)
  209.

\bibitem{1998Cobis}
A.~Cobis, D.~V. Fedorov, A.~S. Jensen, Phys. Rev. C 58 (1998) 1403.

\bibitem{2014Redondo}
C.~Romero-Redondo, S.~Quaglioni, P.~Navr\'atil, H.~G., Phys. Rev. Lett. 113
  (2014) 032503.

\bibitem{2012Bacca}
S.~Bacca, B.~N., A.~Schwenk, Phys. Rev. C 86 (2012) 034321.

\bibitem{2009Forssen}
C.~Forsse\'en, E.~Caurier, P.~Navr\'atil, Phys. Rev. C 79 (2009) 021303(R).

\bibitem{1964Lane}
A.~M. Lane, Nuclear Theory. Pairing Force Correlations and Collective Motion,
  New York, Benjamin, 1964.

\bibitem{2003Dean}
D.~J. Dean, M.~Hjorth-Jensen, Rev. Mod. Phys. 75 (2003) 607.

\bibitem{2001Barranco}
F.~Barranco, P.~F. Bortignon, R.~A. Broglia, E.~Col\`{o}, G.~Vigezzi, European
  Physical Journal A 11 (2001) 385.

\bibitem{2010Potel}
G.~Potel, F.~Barranco, E.~Vigezzi, R.~A. Broglia, Phys. Rev. Lett. 105 (2010)
  172502.

\bibitem{2008Tanihata}
I.~Tanihata, M.~Alcorta, D.~Bandyopadhyay, et~al., Phys. Rev. Lett. 100 (2008)
  192502.

\bibitem{2007Myo}
T.~Myo, K.~Kat\={o}, H.~Toki, K.~Ikeda, Phys. Rev. C 76 (2007) 024305.

\bibitem{2006Nakamura}
T.~Nakamura, A.~M. Vinodkumar, T.~Sugimoto, et~al., Phys. Rev. Lett. 96 (2006)
  252502.

\bibitem{1937Beth}
E.~Beth, G.~Uhlenbeck, Physica 4 (1937) 915.

\bibitem{1998Kruppa}
A.~T. Kruppa, Physcis Letters B 431 (1998) 273.

\bibitem{2010Ono}
T.~Ono, Y.~R. Shimizu, N.~Tajima, S.~Takahara, Phys. Rev. C 82 (2010) 034310.

\bibitem{2012npaIdBetan}
R.~M. Id~Betan, Nuclear Physics A 879 (2012) 14.

\bibitem{2012prcIdBetan}
R.~Id~Betan, Phys. Rev. C 85 (2012) 064309.

\bibitem{1980Maier}
C.~H. Maier, L.~S. Cederbaum, W.~Domcke, J. Phys. B 13 (1980) L119.

\bibitem{1956Cooper}
L.~N. Cooper, Phys. Rev. 104 (1956) 1189.

\bibitem{1980Lawson}
R.~D. Lawson, Theory of the nuclear shell model, Clarendon Press, Oxford, 1980.

\bibitem{2008IdBetan}
R.~M. Id~Betan, G.~G. Dussel, R.~J. Liotta, Phys. Rev. C 78 (2008) 044325.

\bibitem{1984Bonche}
P.~Bonche, S.~Levit, D.~Vautherin, Nuclear Physics A 427 (1984) 278.

\bibitem{2005IdBetan}
R.~Id~Betan, R.~J. Liotta, N.~Sandulescu, T.~Vertse, R.~Wyss, Phys. Rev. C 72
  (2005) 054322.

\bibitem{1982Vertse}
T.~Vertse, K.~F. Pal, Balogh, Computer Physics Communications 27 (1982) 309.

\bibitem{1994Thompson}
I.~J. Thompson, M.~V. Zhukov, Phys. Rev. C 49 (1994) 1904.

\bibitem{1986Bohm}
V.~Bohm, Quantum Mechanics: Foundations and Applications, Springer-Verlag, New
  York., 1986.

\bibitem{nndc}
\href {http://arxiv.org/abs/http://www.nndc.gov/chart}
  {\path{arXiv:http://www.nndc.gov/chart}}.

\bibitem{tunl}
\href {http://arxiv.org/abs/http://www.tunl.duke.edu/nucldata}
  {\path{arXiv:http://www.tunl.duke.edu/nucldata}}.

\bibitem{1989Slaus}
I.~Slaus, Y.~Akaishi, H.~Tanaka, Physics Reports 173 (1989) 257.

\bibitem{1995Ixaru}
L.~G. Ixaru, M.~Rizea, V.~T., Computer Physics Communications 85 (1995) 217.

\bibitem{1996Liotta}
R.~J. Liotta, E.~Maglione, N.~Sandulescu, T.~Vertse, Physics Letters B 367
  (1996) 1.

\bibitem{1972Migdal}
A.~B. Migdal, A.~M. Perelomov, V.~S. Popov, Sov. J. Nucl. Phys. 14 (1972) 488.

\bibitem{2012Wang}
M.~Wang, G.~Audi, A.~H. Wapstra, F.~G. Kondev, M.~MacCormick, X.~Xu,
  B.~Pfeiffer, Chinese Physics C 36 (2012) 1603.

\bibitem{1997Aoi}
N.~Aoi, K.~Yoneda, H.~Miyatake, Y.~Ogawa, H.~Yamamoto, T.~Ideguchi, E.~Kishida,
  et~al., Nucl. Phys. A 616 (1997) 181.

\bibitem{2009Hagino}
K.~Hagino, H.~Sagawa, T.~Nakamura, S.~Shimoura, Phys. Rev. C 80 (2009)
  031301(R).

\bibitem{1968Schiff}
L.~I. Schiff, Quantum Mechanics, McGraw-Hill Book Company. New Yourk, 1968.

\bibitem{2002Carvalho}
C.~A.~A. de~Carvalho, H.~M. Nussenzveig, Phys. Rep. 364 (2002) 83.

\bibitem{1982Newton}
R.~Newton, Scattering Theory of Waves and Particles, Springer, New York, 1982.

\end{thebibliography}

\end{document}